\documentclass{iaus}
\usepackage{graphicx}

\title[Constraining the Properties of SNe~Ia Progenitors from Light Curves]
{Constraining the Properties of SNe~Ia Progenitors from Light Curves}
\author[B. Sadler et al.]
{B. Sadler$^1$, Peter Hoeflich$^1$, E. Baron$^2$, K. Krisciunas$^{3}$, G. Folatelli$^{4}$, M. Hamuy$^{4}$,  M. Khokhlov$^{5}$, M. Phillips$^{6}$, N. Suntzeff$^{3}$, L. Wang$^{3}$}

\affiliation{$^1$Dept. of Physics, Florida State University,USA \\[\affilskip]
$^2$ Dept. of Physics and Astronomy, University of Oklahoma, \\[\affilskip]
$^3$ Dept. of Physics, Texas A \& M University,  \\[\affilskip]
$^4$ Dept. de Astronomia, Universidad de Chile, Santiago, Chile, \\[\affilskip]
$^5$ Dept. Astronomy and Astrophysics, University of Chicago, USA, \\[\affilskip]
$^6$ Las Campanas Observatory, La Serena, Chile}

\pubyear{2008}
\volume{xxx}  
\pagerange{119--126}
\jname{Title of your IAU Symposium}
\editors{A.C. Editor, B.D. Editor \& C.E. Editor, eds.}
\begin{document}

\maketitle

\begin{abstract}
We present an analysis of high precision V light curves (LC) for 18 local Type Ia Supernovae, SNe~Ia, as obtained with the same telescope and setup at the Las Campanas Observatory (LCO).
This homogeneity provides 
an intrinsic accuracy a few hundreds of a magnitude both with respect to individual LCs and between different objects. Based on the Single Degenerate Scenario, SD,
we identify patterns which have been predicted by model calculations as signatures of the progenitor and accretion rate which change the explosion energy 
and the amount of electron capture, respectively.  Using these templates as principle components and the overdetermined system of SN pairs, we  reconstruct
the properties of progenitors and progenitor systems. All LCO SNe~Ia follow the brightness decline relation but 2001ay. After subtraction of
the two components, the remaining scatter is reduced to $\approx 0.01 $ to $0.03^m$.
 Type  SNe~Ia seem to originate from progenitors with Main Sequence masses, $M_{MS} > 3~M_\odot$ with the exception of two subluminous SNe~Ia with $M_{MS} < 2 M_\odot$.
 The component analysis indicates a wide range of accretion rates in the progenitor systems closing the gap to accretion induced collapses (AIC).
 SN1991t-like objects show differences in $dm15$ but no tracers of our secondary parameters. This may point
to a different origin such as DD-Scenario or the Pulsating Delayed Detonations. SN2001ay does not follow the
decline relation. It can be understood in the framework of C-rich WDs, and this group may produce an anti-Phillips relation.
 We suggest that this may be a result of a common envelope phase and mixing during central He burning as in SN1987A. 
\end{abstract}

\firstsection 
\section{Introduction}

\begin{figure}[t]
\begin{center}
\includegraphics[width=3.4in]{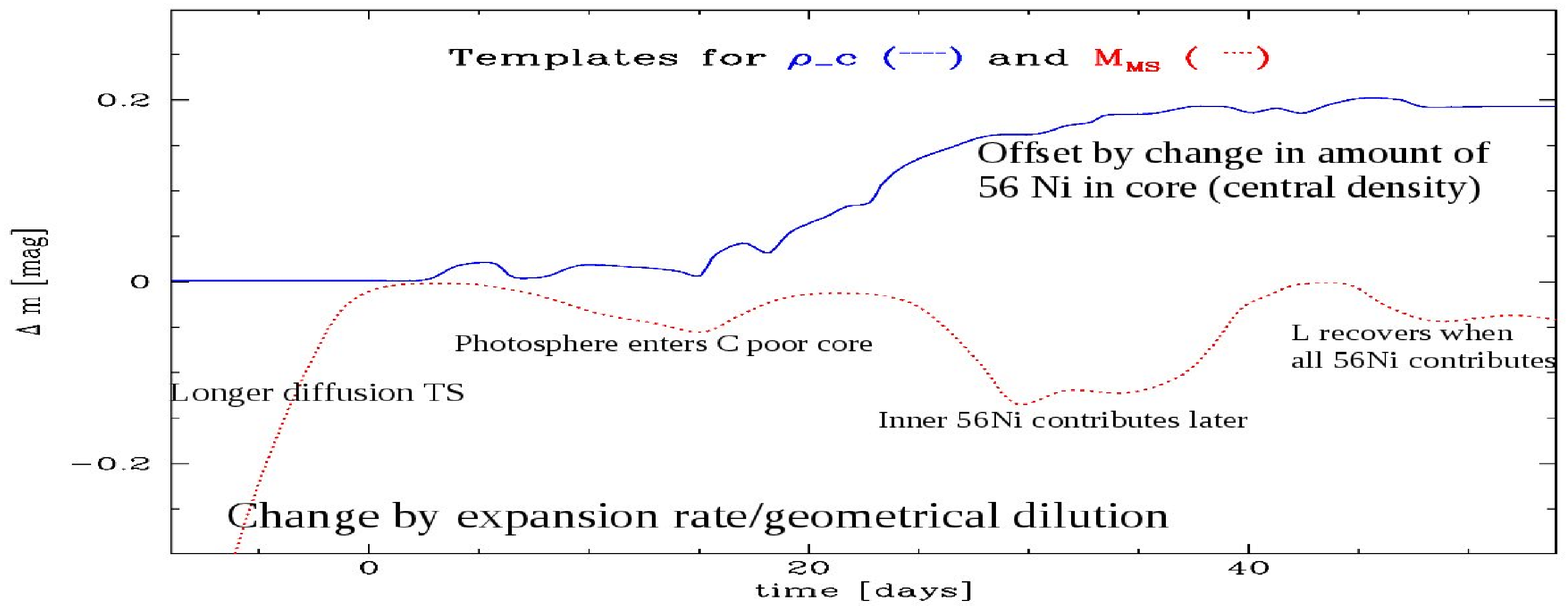}
 \caption{Principle components based on theoretical models. We show the
difference in V brightness $\Delta m(t)$ (in magnitudes) as a function of time (in days) relative to a
reference model with solar metallicity and a main sequence mass $M_{MS} $ of $7 M_\odot$ and
a central density of the exploding WD of $\rho_c$ of $2 \times 10^9 g~cm^{-3}$. The plots are for models with  $7 M_\odot$
and $5 \times 10^{9} g~cm^{-3}$, respectively. The annotation on the graphs give the main reason for the differences.
}
\label{templates}
\end{center}
\end{figure}

SNe~Ia are thermonuclear explosions of WD (\cite{hf60}) i.e. end-stages of stellar evolution of stars between 1 and 7 $M_\odot$.
Most likely, they result from the explosion  of a C/O-WD with a mass close to the Chandrasekhar limit ($M_{Ch}$), which accretes matter through Roche-lobe overflow
in a  single degenerate scenario (SD) (\cite{whelan73}), or merging of two degenerate WDs (DD) (\cite{webbink84,it84}).
 We regard SDs as most likely for the majority of SNe~Ia because e.g. of the homogeneity in LCs and spectra,
though there is strong evidence of contributions of both to the SNe~Ia population (see \cite{hoeflich06},
and references therein). One of the keys is the empirical relation between maximum brightness and the rate of decline, $dm15$,(\cite{p93}).
 From theory, $dm15$ is well understood: LCs are powered by radioactive decay of $^{56}Ni$ (\cite{colgate69}).
  More $^{56}Ni$ increases the luminosity
and causes the envelopes to be hotter. Higher temperature means higher opacity and, thus, longer diffusion time scales
and, thus, slower decline rates after maximum light (\cite{hoeflich96a,nugent97,maeda03,kasen10}). The existence of a
$dm15$-relation holds up for virtually all scenarios as long as there is an excess amount of stored energy to be released
(\cite{hoeflich96a}).

  Within SDs, the favorite models are $M_{Ch}$ explosions
in which burning starts as deflagration which, at some point, transitions to a detonation DDT (\cite{k91}). The
 $^{56}Ni$ production depends mostly on the DDT, i.e. one parameter. The expanding SN-envelopes have
 similar velocity and density structures because  same masses,
most of the WD undergoes burning, and the nuclear binding energies of the burning products are nearly the same.
 A dispersion of $0.2...0.3^m$ is expected (see Fig. \ref{templates}). Its origin 
can be related to the progenitor and accretion from the donor star: Changes in the metallicity and the Main Sequence Mass, $M_{MS}$, will change the size of the explosion energies
 by about 20 \% because their influence of the size of C-depleted core formed during stellar He-core burning (\cite{hwt98,dominguez00}). Increasing accretion rates
 will decrease the central WD density  $\rho_c$ at the time of explosion and, consequently, an increased electron
 capture will produce more  stable iron-group elements at the expense of $^{56}Ni$.
High precision LCs have been obtained at LCO
for 18 local SNe~Ia (\cite{contr10,fol10}) and the individual, theoretical signatures have been recovered
( \cite{h10}).
Note that a similar pattern as by $\rho_c$  can be produced by off-center $^{56}Ni$ distributions
but $\Delta m$ increases later at  $\approx $ 60 to 100 days.

\begin{figure}[h]
\begin{center}
\includegraphics[width=3.4in]{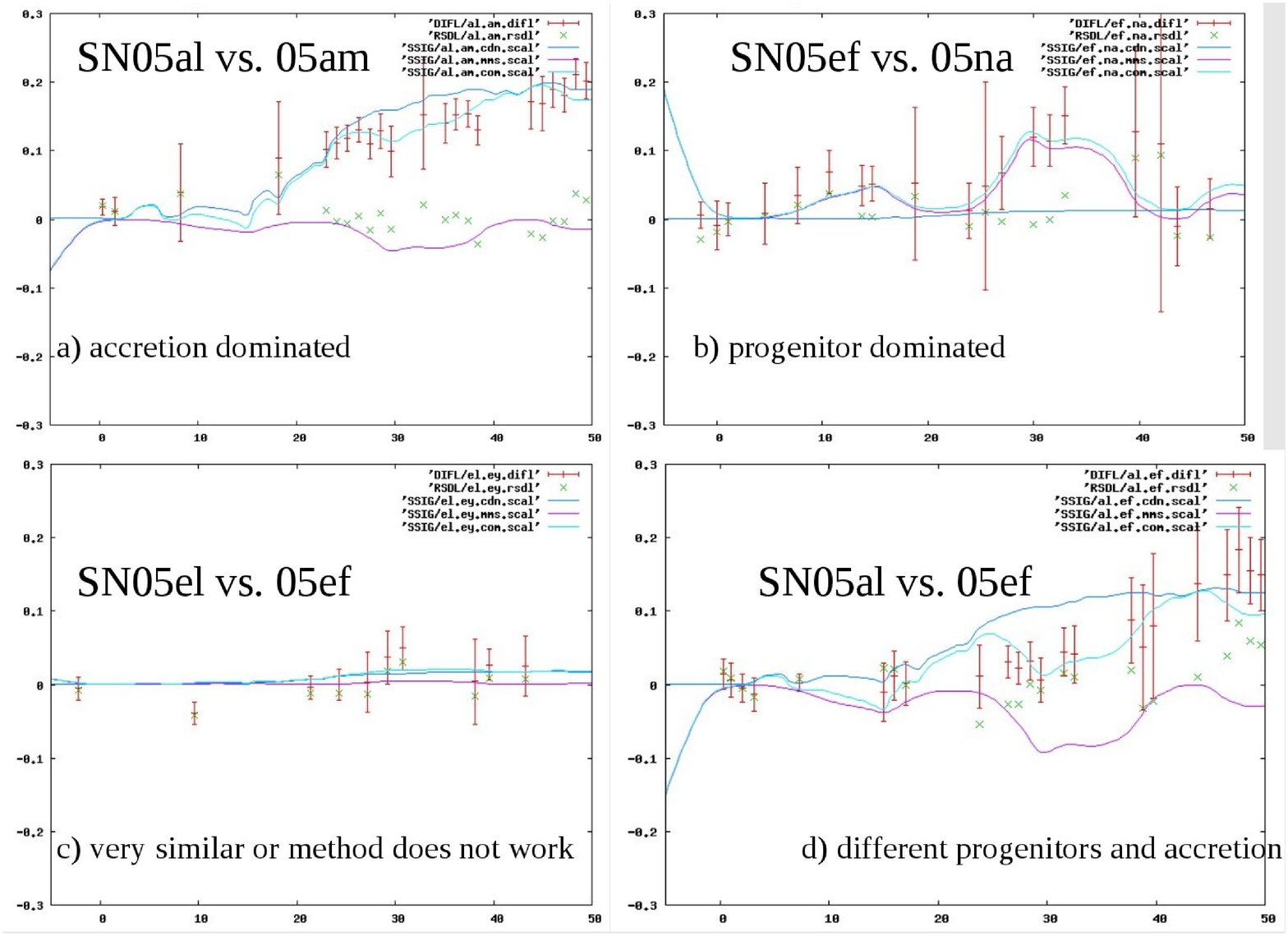}
 \caption{Best fits to the observations of individual pairs of SNe~Ia. We give the weighted components,
their sums along the observations (with error bars) and the residuals (crosses) for cases which are dominated
by the central density/accretion rate (upper left), in the progenitors (upper right), similar $\rho_c$ and $M_\odot$ (lower left),
and a mixed case (lower right). Note that the residuals are small and, within the error bars, consistent with zero.}
\label{res}
\end{center}
\end{figure}
\section{Diversity of Type Ia Supernovae}

\begin{figure}[t]
\begin{center}
\includegraphics[width=3.0in]{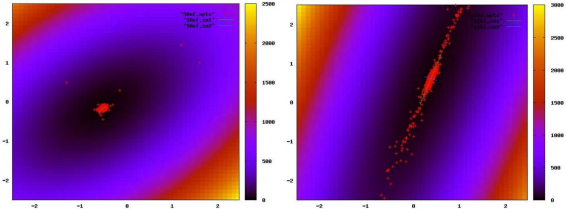}
\caption{Probability distributions of $\tilde{\lambda}_{ij}$ for the pairs of SN2005al/SN2005am and SN2005el/sn2005ef
based on MC solutions for the overdetermined system. Sparse time coverage produce large uncertainties in the eigenvalues.}
\label{error}
\end{center}
\end{figure}

{\bf Secondary parameters:}
 Two SNe~Ia may differ in both $\rho_c$/accretion and the C-depleted core/$M_{MS}$. Therefore, we employed component analyses
to study secondary parameters. We use V because this color is hardly effected by metallicity, asphericity effects, and k-corrections.
 The differences $\Delta m(t)$ in LC pairs are described by
$$\Delta m_{ij,obs}(t) =  \sum _{k=1,2} \lambda_{k}(ij) f_k(t) + Res_{ij}(t)$$
with $\lambda_k(ij)$ being the coefficients for a pair of SN $i$ and $j$ , $f_k(t) $ the principle components,
and $ Res(t)$ the residuals. In our sample, we include 18 SN from LCO, i.e. 153 pairs. Only 36 $\lambda_{ij}=g_k(i)/g_k(j)$'s
are independent where $g_k(i)$ is the eigenvalue of $f_k$ to be attributed a specific SNe~Ia. By solving the overdetermined
system for $g_i$ using a Simplex Method (\cite{nelder65}),
 we obtain most likely values for $\tilde{\lambda}_k(ij)$.
 In Fig. \ref{res}, the pairs SN2005al/SN2005am, SN2005el/sn2005ef, SN~2005ef/SN~2005na and SN~2005al/SN2005ef
are given for $\tilde \lambda_(ij)$. Overall, the residuals are consistent with zero but fits are not unique due to
errors in brightness and time coverage (see Fig. \ref{error}).
 Remaping the individual $g_k(i)$ to $M_{MS}$ and $\rho_c$, shows that (a)
$\rho_c$ are evenly distributed from  $1 \times 10^9 g~cm^{-3} $ to $7 \times 10^9 g cm^{-3}$, i.e. close densities leading to
an accretion induced collapse (AIC) (b) SN~Ia come from massive progenitors with $M \geq 3 M_\odot$ but the two subluminous
SNe~Ia indicate $M_{MS}$ between 1 ... 2 $M_\odot$.
All pairs of SN1991t-like objects show $\lambda_{k}(ij)=0$ in all components. Either, they are very similar despite
differences in $dm15$, or they lack a central region of high densities and similar  C/O in the progenitors as may
be expected for mergers (DD) or pulsating delayed detonation models.  For more details and a complete analysis of all
28 LCO SNe~Ia, see Sadler et al. (in preparation).

\begin{figure}[t]
\begin{center}
\includegraphics[width=3.4in]{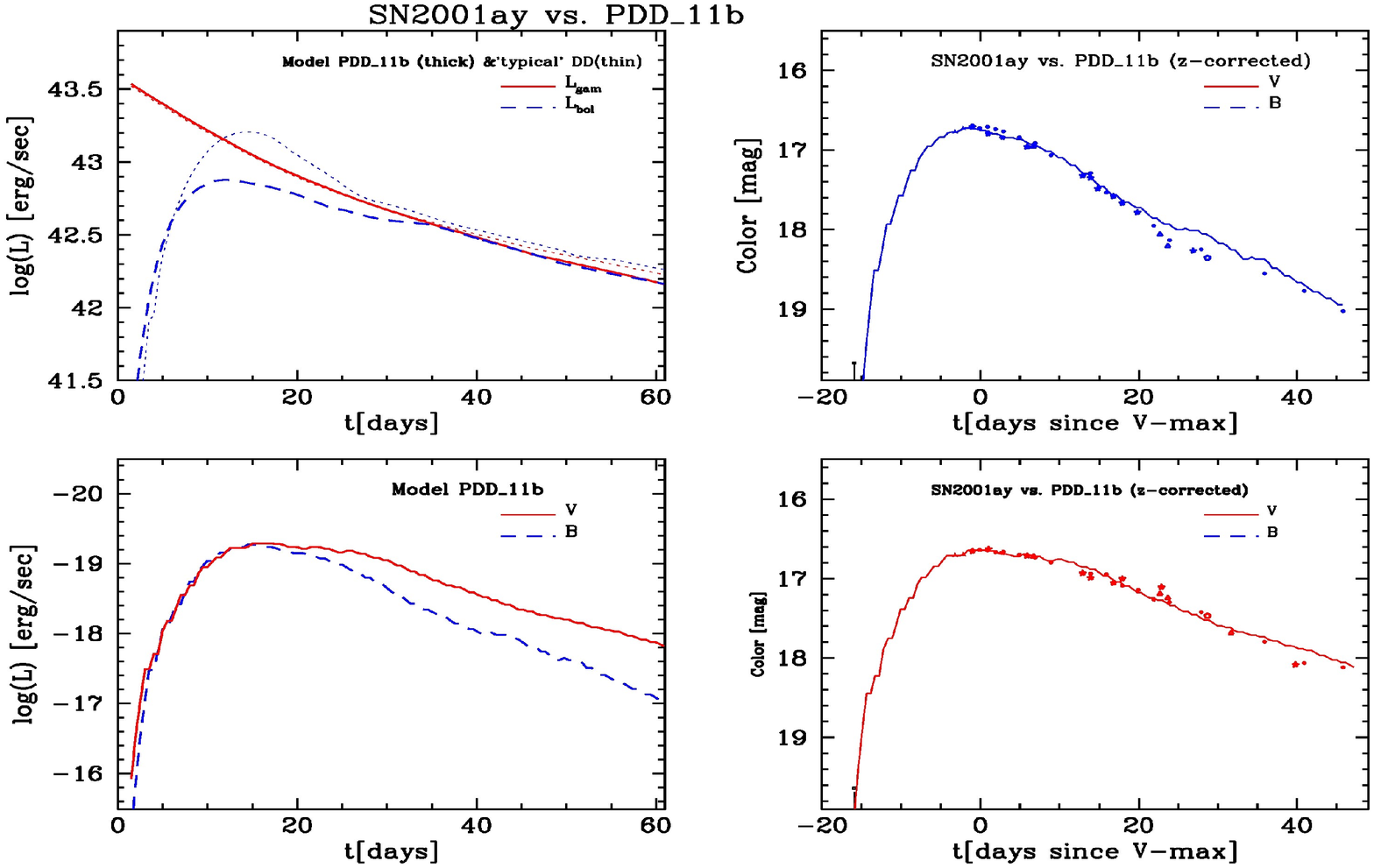}
\caption{
B and V LCs of SN2001ay (Krisciunas et al.,2011) in comparison with theory. We give the
instantaneous deposition by radioactive matter for PDD11b, a pulsating model
with a carbon rich core, and a classical DD model (upper left), and the
B and V (lower left). The DD model has been scaled by 0.037 dex to
equal the gamma-ray input at maximum light.
  The comparison between SN2001ay and PDD11b in B(upper right)
and V (lower right) as a function of time since maximum light in  V observed. We assume a
distance module  m-M of $35.75^m$, and a total reddening  E(B-V) of $0.06^m$.}
\label{sn2001ay}
\end{center}
\end{figure}
\noindent{\bf SN2001ay} shows that nature is even more diverse. SN2001ay is slower than any SNe~Ia known, combined with a fast rise of some 16 days
 (Krisciuanas et al. 2011). SN2001ay would be brighter by about $1^m$ based on $dm15$ and the distance to the host galaxy.
 In fact, $dm15$  is slower than implied by the instantaneous energy input by radioactive decays $\dot{E}_\gamma$.
 We submit that, still, this SN can be understood within the physics underlying the dm15 relation, and in the framework of pulsating delayed detonation models originating from
a $M_{Ch}$ mass WD but with a core of some 80\% C rather instead of the 15 to 20 \% usual for stellar central He burning.
Higher C fraction means more nuclear energy by $^{12}C(\alpha,\gamma)^{16}O$ by $\approx $ 40 \% and faster expansion of the inner layers.
 Faster expansion means that a larger fraction of the energy by $^{56}Ni$ decay goes into expansion work rather than boosting 
the luminosity at maximum light, lower optical depth and shorter rise time. In our models,
the the maximum brightness is smaller than the instant radioactive energy release $\dot{E}\gamma$ (\cite{arnett90}), and the light
curves approaches $\dot{E}\gamma$ "from below". Our model agrees reasonably well with the observations (Fig. \ref{sn2001ay}).
 The reason for a high $C$ abundance we can only speculate. During the early stages of central He burning,
 high C abundances are produced by $^4He(2 \alpha,\gamma)^{12}C$ burning but, at the end,  $^{12}C(\alpha, \gamma)^{16}O $
depletes $^{12}C$ unless strong mixing of $He$ avoids this phase. A possible path may be a common envelope like in the progenitor of SN1987A.
  For more details, see Baron et al. (in preparation).

{}

\end{document}